\documentclass[twocolumn,showpacs,aps,floatfix]{revtex4}
\usepackage{amsmath2000}
\usepackage{graphicx}
\usepackage{dcolumn}
\usepackage{bm}

\begin{document}

\title{Ab initio oscillator strengths for transitions between J=1 odd and J=1,2
even excited states of Ne I }
\author{I. M. Savukov}
 \email{isavukov@princeton.edu}
 \homepage{http://www.princeton.edu/~isavukov}

\affiliation{Department of Physics, Princeton University,
Princeton, New Jersey 08544}

\date{\today}
\begin{abstract}
{\it Ab initio} theory is developed for radiative transitions
between excited states of neon. Calculations of energies for even
excited states J=1, J=2 supplement our previous calculations for
J=1 odd excited states. Line strengths for transitions between J=1
odd and J=1,2 even states of Ne I are evaluated. A comparison with
experiments and semiempirical calculations is given.
\end{abstract}

\maketitle

\section{Introduction}

Development of {\it ab initio} theories for neutral open-shell
atoms is a difficult task since the interaction between electrons
of an open shell is strong and cannot be treated perturbatively.
Nevertheless, some progress in two-valence and even three-valence
electron atoms has been achieved with the combination of
configuration-interaction (CI) method and many-body perturbation
theory (MBPT)~\cite{cimbpt:96a,isav5,disser}. Particle-hole states
of closed-shell atoms have additional difficulty that conventional
perturbation theory does not converge for hole states. We solved
the convergence problem by modifying
denominators~\cite{neonepr,disser}. As a result, we were able to
achieve agreement with experiment for neon energies of J=1 odd
excited states and oscillator strengths (averaged over many
measurements) of transitions to the ground state. In this paper,
we extend our application of the CI+MBPT method to neon transition
between excited states. If we succeed, our understanding of neon
atom and low-$Z$ neonlike ions will be substantially improved

Fairly accurate (about 5\%) measurements of many transition rates
between excited states are available, providing important tests of
theory. In addition, semiempirical calculations can be compared
with our calculations.
 For example,
many transition rates along the neon isoelectronic sequence were
calculated  by~\citet{neon:93} with a general
configuration-interaction code (CIV3)~\cite{neon:75}. In
calculations a few parameters were adjusted to fit experimental
energies. However, even after such adjustments, the results still
disagreed significantly with other semiempirical calculations
by~\citet{neon:98a} and experiments. The latter theory was more
successful, giving results in close agreement with experiments.
  No pure {\it ab initio} theory, as far as we know, was (successfully) applied
previously to calculations of transitions between neon excited
states. For transitions to the ground state, elaborate {\it ab
initio} calculations exist (\citet{neon:98}), but agreement with
experiment for an oscillator strength of the $\left[
2p_{3/2}^{-1}3s_{1/2}\right] _{1}$ neon state is unsatisfactory.
These calculations are more effective in heavier noble-gas atoms,
where the agreement with experiment is achieved for Ar, Kr, and Xe
atoms. In neon, on the other hand, our new calculations with
CI+MBPT method, give results close to the average experimental
values.

It is a well-known fact that neon transitions between excited
states are sensitive to the accuracy of fine-structure splittings.
Semiempirical theories avoid this difficulty by introducing and
adjusting several parameters to match energies of multiplets as
precisely as possible. For example, using quantum defect method,
~\citet{neon:98a} was able to obtain very small root-mean square
deviations for energies. As a result, he also was able to obtain
transition oscillator strengths that agree well with experiment.
Getting accurate fine structure intervals without parametric
adjustments is a challenging task. We will demonstrate in this
paper that CI calculations corrected with second-order MBPT give
energies and oscillator strengths for transitions between excited
states with the precision comparable to the precision of best
semiempirical calculations.

 The transition data in neon and other noble gases
are needed for plasma physics and studies of discharges with many
industrial applications in lamps and gas lasers. The opacity
project~\cite{opac} is another motivation behind many calculations
(one example is given in Ref.~\cite{opac2}) in neon-like ions.
Understanding of neon atom can be beneficial for the development
of atomic structure methods which are needed for many
applications. One important application of atomic structure is
calculations of parity-nonconservation amplitudes in heavy atoms
with one or a few valence electrons which require a clear
understanding of correlation effects in these atoms.
One-valence-electron MBPT has similar convergence problems to the
hole MBPT after a core is excited. Modification of denominators
according to our prescription might be one key to the solution of
a puzzling problem that third-order energy in Cs agrees worse with
experiment than second-order energy. Calculations of electron
dipole moments in particle-hole atoms is another, more direct
application of our particle-hole theory. Furthermore, the CI+MBPT
method and convergent hole perturbation theory can be generalized
for more complicated atoms with more than one particle or hole and
properties of these atoms can be explored beyond Hartree-Fock
approximation.

In this paper, first we will briefly describe our method of
calculations (more details are given in
Ref.~\cite{neonepr,disser}); then, we will compare CI+MBPT and
experimental energies for J=1 and J=2 even states. This comparison
gives an estimate on accuracy of our wave functions. Next, we will
show our results for transition line strengths. Finally, we
compare our theory with other semiempirical calculations and
experiments.

\section{CI+MBPT calculations}
\subsection{Energies and oscillator strengths for J=1 odd neon states}
 The Rayleigh-Schr\"{o}dinger variant of
second-order MBPT, given in Ref.~\cite{neon:01},
 has low accuracy
for neon (does not improve lowest order approximation), therefore
we developed a fast convergent variant of MBPT
~\cite{neonepr,disser}. This perturbation theory can be understood
from couple-cluster single-double equations~\cite{neon:95}. Simply
put, we modify some denominators in the perturbation terms to take
into account the strong interaction between a hole and a core
electron or between core electrons non-perurbatively. The
advantage of this approach compared to the couple-cluster method
of~\cite{neon:95} is simplicity and speed of calculations. With
our fast convergent MBPT method, we are able to improve the
accuracy of hole energies and fine-structure splittings of light
neonlike ions already after adding second-order MBPT corrections.
Apart from Coulomb correlation corrections, the Breit magnetic
interaction is also included, but small frequency-dependent Breit,
quantum-electrodynamic, reduced-mass, and mass-polarization
corrections are omitted (the analysis of these small corrections
is given in Ref.~\cite{neon:95}.

To calculate particle-hole energies, we construct a model CI
space~\cite{neon:01}, compute effective Hamiltonian in this space,
which also includes second-order MBPT corrections, and solve an
eigen value problem. Along with energies we obtain wave functions,
which were used to calculate oscillator strengths for transitions
to the ground states~\cite{neonepr,disser}. The energies of neon
particle-hole J=1 odd states and oscillator strengths were in very
good agreement with experiment after using a relatively small CI
space, 52. Pure {\it ab initio} energies differed from
experimental energies by 0.0069 a.u., but after subtraction of the
systematic shift (which does not make much difference in
transition calculations), the agreement was improved to the level
of 0.0001 a.u.  for almost all states. We will use the same wave
functions for J=1 odd states in our calculations of transitions
from J=1 odd to J=1 and J=2 even states. The accuracy of energies
of even states involved in the transitions will be illustrated
below.

\subsection{Energies for J=2 and J=1 even neon states}
\begin{table}
\caption{Calculations of neon energy levels for J=1 even states.
 In the third and forth columns
``CI-8'' and ``CI-32'' mean that in our calculations the size of
the CI matrices were 8 and 32, respectively. $\Delta $ is the
difference between theoretical and experimental energies. All
energies are in atomic units.} \label{Netab1}
\begin{center}
\begin{tabular}{cccccc}
\hline\hline
J=1 even  & NIST & CI-8    & CI-32  & $\Delta $ & shifted $\Delta $ \\
\hline
p$_{3/2}^{-1}$3p & 0.67551 & 0.6687 & 0.6690 & 0.0065 & 0.0005 \\
p$_{3/2}^{-1}$3p & 0.68400 & 0.6789 & 0.6787 & 0.0053 & -0.0007 \\
p$_{1/2}^{-1}$3p & 0.68696 & 0.6820 & 0.6817 & 0.0052 & -0.0008 \\
p$_{1/2}^{-1}$3p & 0.68818 & 0.6833 & 0.6830 & 0.0052 & -0.0008 \\
p$_{3/2}^{-1}$4p & 0.74048 & 0.7337 & 0.7339 & 0.0066 & 0.0006 \\
p$_{3/2}^{-1}$4p & 0.74274 & 0.7370 & 0.7367 & 0.0060 & 0.0000 \\
p$_{1/2}^{-1}$4p & 0.74567 & 0.7398 & 0.7396 & 0.0061 & 0.0001 \\
p$_{1/2}^{-1}$4p & 0.74590 & 0.7402 & 0.7399 & 0.0060 & 0.0000 \\
\hline
\end{tabular}
\end{center}
\end{table}
\begin{table}
\caption{Calculations of neon energy levels for J=2 even states.
The size of the CI matrix is 32. $\Delta $ is the difference
between theoretical and experimental energies. All energies in
atomic units} \label{Netab2}
\begin{center}
\begin{tabular}{ccccc}
\hline\hline J=2 even  & NIST & CI-32 & $\Delta $ & shifted
$\Delta $ \\ \hline
p$_{3/2}^{-1}$3p & 0.68265 & 0.6775 & 0.0051 & -0.0007 \\
p$_{3/2}^{-1}$3p & 0.68489 & 0.6800 & 0.00487 & -0.0009 \\
p$_{1/2}^{-1}$3p & 0.68736 & 0.6825 & 0.00485 & -0.0010 \\
p$_{3/2}^{-1}$4p & 0.74222 & 0.7363 & 0.00588 & 0.0001 \\
p$_{3/2}^{-1}$4p & 0.74285 & 0.7372 & 0.00565 & -0.0001 \\
p$_{1/2}^{-1}$4p & 0.74591 & 0.7401 & 0.00578 & -0.0000 \\ \hline
\end{tabular}
\end{center}
\end{table}

In this section we present our calculations of energies for even
states. We use the same formalism and numerical method as in
Ref.~\cite{neonepr,disser}. The spline cavity is chosen 80 a.u.,
the number of splines is chosen 40, and the maximum orbital
momentum is chosen 5. For excited states, a $V^{N-1}$ Hartree-Fock
(HF) potential basis (see ~\cite{disser}) is built from HF $V^{N}$
spline basis by diagonalization of a one-electron Hamiltonian to
take into account the major part of the interaction of an excited
electron with a hole. Such procedure speeds up the convergence of
CI and reduces uncertainty in denominators of a perturbation
theory. A $V^{N-1}$ HF basis used in Ref.~\cite{neon:95} was
constructed by solving differential HF equations and gives similar
results, but our basis does not require rewriting HF code and
therefore is more convenient. The calculations of energies are
shown in two tables.

In Table~\ref{Netab1} we compare with experiment our theoretical
energies of  J=1 even states. Energies calculated in a model space
of size 32 agree better with experiment than energies calculated
in smaller size (8) model space. CI space 32 can be considered as
optimal since a larger number of configurations does not improve
much accuracy. Note that the size of optimal CI space depends on
the choice of a starting potential since inadequate initial
approximation is corrected by diagonalization of the CI matrix. An
1\% systematic shift is present, which can be attributed to
inaccuracy of hole energies; however this shift does not affect
much accuracy of transition rates and is relatively unimportant.
More important for weaker transitions is the fact that after
subtracting this shift, we obtain very small residual deviations,
which bode well for the accuracy of singlet-triplet mixing
coefficients and transition amplitudes.

Similar agreement of energies is obtained for J=2 even states in
Table~\ref{Netab2}. Again, we have almost the same systematic
shift, and after its subtraction, only a very small difference
between experimental and theoretical energies remains. Therefore,
we have reason to expect good precision for transitions between
the excited states which are considered next.

\subsection{Transitions between neon excited states}
\begin{table}
\caption{Transitions between neon excited states, from J=1 odd to
J=1 even. In notations, p$_{3/2}^{-1}$ and p$_{1/2}^{-1}$ are hole
states, and particle states are standing immediately to the right.
For unique specification, experimental wavelengths are also
provided. The sizes of model space for even states are given, but
for J=1 odd states the size is 50, the same in all cases. The
experimental NIST and theoretical line strengths (in columns
``CI-8'' and ``CI-32'') are expressed in a.u.; $\Delta$ denotes
the relative deviations of the theoretical line strengths from the
experimental line strengths} \label{Netab3}
\begin{center}
\begin{tabular}{llclcccc}
\hline\hline Transition & $\lambda $, \AA & NIST & Ac. & CI-8  & CI-32 & +RPA & $\Delta$\\
 \hline
p$_{3/2}^{-1}$3s-p$_{3/2}^{-1}$3p & 6385 & 12.4 & B- & 12.5 & 12.6 & 12.2 & 2\%\\
p$_{3/2}^{-1}$3s-p$_{1/2}^{-1}$3p & 6032 & 1.82 & B- & 1.83 & 1.88 & 1.81& 1\%\\
p$_{3/2}^{-1}$3s-p$_{1/2}^{-1}$3p & 6130 & 0.23 & B- & 0.32 & 0.23 & 0.22& 4\%\\
p$_{3/2}^{-1}$3s-p$_{3/2}^{-1}$4p & 3502 & 0.076& D  & 0.114& 0.102& 0.084 &11\%\\
p$_{3/2}^{-1}$3s-p$_{3/2}^{-1}$3p & 7247 & 5.27 & B- & 5.99 & 5.93 & 5.70 &8\%\\
p$_{1/2}^{-1}$3s-p$_{3/2}^{-1}$3p & 8085 & 0.094& B  & 0.121& 0.107& 0.103&10\% \\
p$_{1/2}^{-1}$3s-p$_{3/2}^{-1}$4p & 3687 & 0.029& D  & 0.038&0.040 &0.032&10\%\\
p$_{1/2}^{-1}$3s-p$_{1/2}^{-1}$3p & 6601 & 9.88 & B- & 9.16 &9.25  & 8.92&10\%\\
p$_{1/2}^{-1}$3s-p$_{3/2}^{-1}$3p & 7026 & 0.971& B  & 0.825& 1.01 & 0.982&1\%\\
p$_{1/2}^{-1}$3s-p$_{1/2}^{-1}$3p & 6719 & 9.75 & B- & 11.0 & 10.9 &10.5&8\%\\
p$_{1/2}^{-1}$3s-p$_{1/2}^{-1}$4p & 3595 & 0.045& D  & 0.070& 0.065&0.053&18\% \\
p$_{1/2}^{-1}$3s-p$_{1/2}^{-1}$4p & 3601 &0.030 & D  &0.050 &0.045 &0.036&21\%\\
 \hline
\end{tabular}
\end{center}
\end{table}
\begin{table}
\caption{Line strengths (a.u.) for transitions between neon
excited states from J=1 odd to J=2 even. Abbreviations are the
same as in Table~\ref{Netab3}} \label{Netab4}.
\begin{center}
\begin{tabular}{llclccc}
\hline\hline Transitions & $\lambda $, \AA & NIST & Acc. & CI-32
&+RPA& $\Delta $
\\ \hline
p$_{3/2}^{-1}$3s-p$_{1/2}^{-1}$3p & 6097.8507 & 10.1 & C+ & 10.2 &9.76& 3\% \\
p$_{3/2}^{-1}$3s-p$_{3/2}^{-1}$3p & 6508.3259 & 20.4 & B- & 21.1 &20.2& 1\% \\
p$_{3/2}^{-1}$3s-p$_{3/2}^{-1}$3p & 6306.5325 & 2.57 & B- & 2.54 &2.62& 2\% \\
p$_{3/2}^{-1}$3s-p$_{3/2}^{-1}$4p & 3516.1960 & 0.074 & D & 0.106 &0.077& 4\% \\
p$_{1/2}^{-1}$3s-p$_{3/2}^{-1}$4p & 3702.2783 & 0.028 & D & 0.034&0.029& 4\%\\
p$_{1/2}^{-1}$3s-p$_{1/2}^{-1}$3p & 6680.1202 & 17.1 & C+ & 16.9 &17.1&0\% \\
p$_{1/2}^{-1}$3s-p$_{3/2}^{-1}$3p & 6931.3788 & 14.3 & B- & 16.6 &15.3& 7\% \\
p$_{1/2}^{-1}$3s-p$_{1/2}^{-1}$4p & 3594.5516 & 0.11 & D & 0.14 &0.11& 0\% \\
p$_{1/2}^{-1}$3s-p$_{3/2}^{-1}$3p & 7175.9155 & 2.62 & B- & 2.51 &2.61& 0\% \\
\hline
\end{tabular}
\end{center}
\end{table}
\begin{table}
\caption{Comparison of experimental and theoretical oscillator
strengths of neon. We adopt compact notations of
Ref.~\cite{neon:98a}: $p_i q_i r_i-p_f q_f r_f$, where $p=2\times
j_a$, $q=2\times K$, $r=J$, $i$ stands for initial, and $f$ stands
for final. The hole angular momentum $j_a$ is coupled with the
orbital momentum of the excited electron $l$ to give the angular
momentum $K$ in the intermediate coupling scheme; $K$ is coupled
with a spin of the excited electron to give a total angular
momentum of particle-hole state $J$. For compete specification,
$nl$ quantum numbers are also provided. Brackets denotes powers of
10. } \label{Netab5}
\begin{center}
\begin{tabular}{lllcccc}
\hline\hline
Transitions& $\lambda$, $\AA$& NIST  & Ref.\cite{neonexp}& Current & Ref.\cite{neon:98a} & Ref.\cite{neon:93}\\
\hline
3s-3p   331-132 & 6097    & 5.03[-1]    &4.98[-1] & 4.86[-1]& 4.88[-1]    & 5.49[-1] \\
3s-3p   331-352 & 6508    & 9.52[-1]    &9.46[-1] & 9.43[-1]  &9.27[-1]    & 1.03[\;0] \\
3s-3p   331-332 & 6307    & 1.24[-1]    &1.26[-1] & 1.26[-1] &   1.26[-1] & 1.20[-1] \\
3s-4p   331-352 & 3516    & 6.39[-3]    & -       & 6.65[-3]  &6.22[-3]&   -\\
3s-4p 111-352 &   3702    & 2.30[-3]    & -       & 2.38[-3]   & 2.27[-3] &   - \\
3s-3p 111-132 & 6680      & 7.78[-1]   & 7.71[-1] & 7.78[-1] &7.78[-1]& 8.18[-1] \\
3s-3p 111-332 & 6931    & 6.27[-1]    &6.27[-1]   & 6.71[-1] &   6.34[-1]    & 6.16[-1] \\
3s-4p 111-132 &   3595    & 9.29[-3]    &  -      & 9.29[-3] &8.59[-3] & - \\
3s-3p 111-352 &   7176    & 1.11[-1]    & 1.24[-1]& 1.10[-1] &1.18[-1]    & 1.30[-1]\\
3s-3p   331-331 &   6385  &   5.90[-1]  & 5.81[-1] &5.78[-1]   &   5.81[-1]    &   6.45[-1] \\
3s-3p   331-111 & 6032 &   9.17[-2]    &   8.36[-2]&9.11[-2]&   8.57[-1] & 7.79[-2] \\
3s-3p   331-131 &   6130    & 1.14[-2] & 1.19[-2] & 1.09[-2] & 1.12[-2]    & 2.49[-2] \\
3s-4p 331-331 & 3502       &   6.59[-3] & -       & 7.29[-3] & 6.60[-3] & - \\
3s-3p   331-311 &   7247  & 2.21[-1]   & 2.36[-1] & 2.39[-1]    &2.21[-1]    & 2.46[-1] \\
3s-3p 111-311 &   8085 & 3.53[-3]    & 3.53[-3]    & 3.87[-3]    &3.02[-3]    & 3.60[-3] \\
3s-4p   111-331&   3687    & 2.39[-3] &  -         & 2.64[-3]    & 2.42[-3]    &   - \\
3s-3p 111-111 & 6601    & 4.55[-1]    & 4.41[-1]   & 4.10[-1] &4.42[-1] & - \\
3s-3p 111-331 & 7026    &   4.20[-2]   & 4.35[-2]  &4.25[-2] & 4.43[-2]    & 3.69[-2] \\
3s-3p 111-131 & 6719 & 4.41[-1]  & 4.41[-1]       & 4.76[-1] &4.42[-1]& 3.70[-1] \\
3s-4p 111-111 & 3595    & 3.80[-3]    &  -         & 4.48[-3] &4.32[-3]    &- \\
3s-4p 111-131 &   3601    & 2.53[-3]&     -       &   3.07[-3]    & 2.55[-3] & -\\
\hline
\end{tabular}
\end{center}
\end{table}
Previously, we calculated oscillator strengths for the transitions
to the ground state; the formula for excited state transitions is
different.  The final expression for the coupled reduced matrix
element after angular reduction in j-j relativistic basis has the
following form:
\begin{multline}
\left\langle F\left\| Z_{J}\right\| I\right\rangle =\sqrt{\left[
J_{F}\right] \left[ J_{I}\right] } \;C_F(a'v')C_I(a,v)\times \nonumber \\
\left\{ (-1)^{J+J_{I}+j_{a}+j_{v^{\prime }}}\left\{
\begin{array}{lll}
J_{I} & J & J_{F} \\
j_{v^{\prime }} & j_{a} & j_{v}
\end{array}
\right\} \delta _{a^{\prime }a}\left\langle v^{\prime }\left\|
Z_{J}\right\|
v\right\rangle +\right.   \nonumber \\
\left. (-1)^{J_{F}+j_{a^{\prime }}+j_{v^{\prime }}+1}\left\{
\begin{array}{lll}
J & J_{I} & J_{F} \\
j_{v} & j_{a^{\prime }} & j_{a}
\end{array}
\right\} \delta _{v^{\prime }v}\left\langle a\left\| Z_{J}\right\|
a^{\prime }\right\rangle \right\}
\end{multline}
We use standard notations of relativistic MBPT methods, see for
example~\cite{dan}. Configuration weights of the final $C_F(a'v')$
and the initial $C_I(a,v)$ states are obtained in CI calculations.
Note that this formula is similar to but different from that for
the transitions between two-particle states~\cite{dan} even if we
neglect small hole-hole matrix elements.  However, it is possible
to modify wave functions to use two-particle matrix elements,
which can be convenient if the program for calculations of
two-particle matrix elements are available.

 Tables~\ref{Netab3} and~\ref{Netab4} show our results of calculations for J=1-J=1
 and J=1-J=2 excited state transitions. Many precisely measured (5\% level) neon
transition rates provide an important test of accuracy of our
calculations. In Table~\ref{Netab3}, the calculations are done
with two numbers of configurations to show improvement in
precision for the larger number of configurations. Important
random-phase approximation (RPA) corrections are included by
replacing ``bare'' matrix element with ``dressed'', RPA matrix
elements (the replacement for two-valence electron atoms was
implemented in~\cite{isav5,disser}). Including RPA corrections
needs some care because a hole state present when the transition
between excited states occurs leads to the convergence problem and
low accuracy of the regular RPA corrections. Our standard cure is
to modify denominators by subtracting radial Slatter integral
$R_{0}(abab)$ thus taking into account the monopole interaction of
a hole with a core electron. This modification of denominators
approximately halves RPA corrections. Adding divided by 2 normal
RPA corrections, we estimated the level of these corrections and
found that they are important and improve agreement with
experiment. In neon, the corrections constitute a few percent part
of a total matrix element, but in heavier noble gas atoms, they
are even larger and more important. In the last column, we place
our best values calculated in CI-32 model space with appropriate
modified-denominator RPA corrections.

The deviation from experiment is consistent with the experimental
accuracy (for example, the accuracy of class ``B'' is in the range
5\% and the deviation from theory is of the same magnitude). For
transitions which have experimental accuracy of classes ``B-'' and
``C+'', the theory is as accurate as or even more accurate than
experiment, but  for the class ``D'', the theory is definitely
more accurate. Still the accuracy of theory is not the same for
all transitions, since some suppressed transitions owing to
cancellation could be more sensitive to fine-structure splittings.

Our final table (Table~\ref{Netab5}) contains our best values of
oscillator strengths with the RPA corrections for comparison with
experiments and other theories. Oscillator strengths are
calculated from line strengths $S$ and transition energies
$\omega$ in atomic units.
\begin{equation}\label{eq2}
  f=\frac{2}{3}\omega S
\end{equation}
Due to a large number of measurements and calculations we
restricted ourselves to comparison with results from a few
sources, which contain further references; for example, references
to many experiments and comparison with several measurements are
given by~\citet{neonexp} where the authors also estimated
uncertainties of their experiment to be about 7\% and of the
others shown in their comparison table to be in the range 10-50\%.
The experiments seem to be in good agreement with each other. NIST
data derived from various sources are in close agreement with
values given by~\citet{neonexp}. The agreement with experiment of
the semiempirical theory by~\citet{neon:98a} is similar to the
agreement of our {\it ab initio} theory. Calculations performed
by~\citet{neon:93} agree worse; for example, for the 6130 \AA
~transition, an experimental value is 0.0114 or 0.0119, our value
is 0.0109, but the value in Ref.~\cite{neon:93} is 0.0249. Overall
agreement of theories and experiments is quite normal.

\section{Conclusions}

In this paper, we have applied CI+MBPT theory for particle-hole
states of closed-shell atoms to calculations of transitions
between excited states of neon. A difficulty that the hole energy
has poor convergence is overcome with modifications of
denominators in MBPT. Good precision for particle-hole states is
illustrated for many energy levels of neon. Apart from energies,
our theory is tested in calculations of line strengths. Agreement
with experimental values is achieved.


\begin{thebibliography}{14}
\expandafter\ifx\csname
natexlab\endcsname\relax\def\natexlab#1{#1}\fi
\expandafter\ifx\csname bibnamefont\endcsname\relax
  \def\bibnamefont#1{#1}\fi
\expandafter\ifx\csname bibfnamefont\endcsname\relax
  \def\bibfnamefont#1{#1}\fi
\expandafter\ifx\csname citenamefont\endcsname\relax
  \def\citenamefont#1{#1}\fi
\expandafter\ifx\csname url\endcsname\relax
  \def\url#1{\texttt{#1}}\fi
\expandafter\ifx\csname
urlprefix\endcsname\relax\def\urlprefix{URL }\fi
\providecommand{\bibinfo}[2]{#2}
\providecommand{\eprint}[2][]{\url{#2}}

\bibitem[{\citenamefont{Dzuba et~al.}(1996)\citenamefont{Dzuba, Flambaum, and
  Kozlov}}]{cimbpt:96a}
\bibinfo{author}{\bibfnamefont{V.~A.} \bibnamefont{Dzuba}},
  \bibinfo{author}{\bibfnamefont{V.~V.} \bibnamefont{Flambaum}},
  \bibnamefont{and} \bibinfo{author}{\bibfnamefont{M.~G.}
  \bibnamefont{Kozlov}}, \bibinfo{journal}{Phys.\ Rev. A}
  \textbf{\bibinfo{volume}{54}}, \bibinfo{pages}{3948} (\bibinfo{year}{1996}).

\bibitem[{\citenamefont{Savukov and Johnson}(2002)}]{isav5}
\bibinfo{author}{\bibfnamefont{I.~M.} \bibnamefont{Savukov}} \bibnamefont{and}
  \bibinfo{author}{\bibfnamefont{W.~R.} \bibnamefont{Johnson}},
  \bibinfo{journal}{Phys. Rev. A.} \textbf{\bibinfo{volume}{65}},
  \bibinfo{pages}{042503} (\bibinfo{year}{2002}).

\bibitem[{\citenamefont{Savukov}(2002)}]{disser}
\bibinfo{author}{\bibfnamefont{I.~M.} \bibnamefont{Savukov}}, Ph.D. thesis,
  \bibinfo{school}{University of Notre Dame} (\bibinfo{year}{2002}).

\bibitem[{\citenamefont{Savukov et~al.}(2002)\citenamefont{Savukov, Johnson,
  and Berry}}]{neonepr}
\bibinfo{author}{\bibfnamefont{I.~M.} \bibnamefont{Savukov}},
  \bibinfo{author}{\bibfnamefont{W.~R.} \bibnamefont{Johnson}},
  \bibnamefont{and} \bibinfo{author}{\bibfnamefont{H.~G.} \bibnamefont{Berry}},
  \bibinfo{journal}{Phys. Rev. A} \textbf{\bibinfo{volume}{66}},
  \bibinfo{pages}{052501} (\bibinfo{year}{2002}).

\bibitem[{\citenamefont{Hibbert et~al.}(1993)\citenamefont{Hibbert, Dourneuf,
  and Mohan}}]{neon:93}
\bibinfo{author}{\bibfnamefont{A.}~\bibnamefont{Hibbert}},
  \bibinfo{author}{\bibfnamefont{M.~L.} \bibnamefont{Dourneuf}},
  \bibnamefont{and} \bibinfo{author}{\bibfnamefont{M.}~\bibnamefont{Mohan}},
  \bibinfo{journal}{At. Data} \textbf{\bibinfo{volume}{53}},
  \bibinfo{pages}{23} (\bibinfo{year}{1993}).

\bibitem[{\citenamefont{Hibbert}(1975)}]{neon:75}
\bibinfo{author}{\bibfnamefont{A.}~\bibnamefont{Hibbert}},
  \bibinfo{journal}{Comp. Phys. Commun.} \textbf{\bibinfo{volume}{9}},
  \bibinfo{pages}{141} (\bibinfo{year}{1975}).

\bibitem[{\citenamefont{Seaton}(1998)}]{neon:98a}
\bibinfo{author}{\bibfnamefont{M.~J.} \bibnamefont{Seaton}},
  \bibinfo{journal}{J. Phys. B} \textbf{\bibinfo{volume}{31}},
  \bibinfo{pages}{5315} (\bibinfo{year}{1998}).

\bibitem[{\citenamefont{Avgoustoglou and Beck}(1998)}]{neon:98}
\bibinfo{author}{\bibfnamefont{E.~N.} \bibnamefont{Avgoustoglou}}
  \bibnamefont{and} \bibinfo{author}{\bibfnamefont{D.~R.} \bibnamefont{Beck}},
  \bibinfo{journal}{Phys. Rev. A} \textbf{\bibinfo{volume}{57}},
  \bibinfo{pages}{4286} (\bibinfo{year}{1998}).

\bibitem[{\citenamefont{Seaton}(1987)}]{opac}
\bibinfo{author}{\bibfnamefont{M.~J.} \bibnamefont{Seaton}},
  \bibinfo{journal}{J. Phys.\ B} \textbf{\bibinfo{volume}{20}},
  \bibinfo{pages}{6363} (\bibinfo{year}{1987}).

\bibitem[{\citenamefont{Hibbert and Scott}(1994)}]{opac2}
\bibinfo{author}{\bibfnamefont{A.}~\bibnamefont{Hibbert}} \bibnamefont{and}
  \bibinfo{author}{\bibfnamefont{M.~P.} \bibnamefont{Scott}},
  \bibinfo{journal}{J. Phys. B} \textbf{\bibinfo{volume}{27}},
  \bibinfo{pages}{1315} (\bibinfo{year}{1994}).

\bibitem[{\citenamefont{Safronova et~al.}(2001)\citenamefont{Safronova,
  C.~Namba, Johnson, and Safronova}}]{neon:01}
\bibinfo{author}{\bibfnamefont{U.~I.} \bibnamefont{Safronova}},
  \bibinfo{author}{\bibfnamefont{I.~M.} \bibnamefont{C.~Namba}},
  \bibinfo{author}{\bibfnamefont{W.~R.} \bibnamefont{Johnson}},
  \bibnamefont{and} \bibinfo{author}{\bibfnamefont{M.~S.}
  \bibnamefont{Safronova}}, \bibinfo{journal}{Natl. Inst. Fusion
  Science-DATA-61} \textbf{\bibinfo{volume}{61}}, \bibinfo{pages}{1}
  (\bibinfo{year}{2001}).

\bibitem[{\citenamefont{Avgoustoglou et~al.}(1995)\citenamefont{Avgoustoglou,
  Johnson, Liu, and Sapirstein}}]{neon:95}
\bibinfo{author}{\bibfnamefont{E.}~\bibnamefont{Avgoustoglou}},
  \bibinfo{author}{\bibfnamefont{W.~R.} \bibnamefont{Johnson}},
  \bibinfo{author}{\bibfnamefont{Z.~W.} \bibnamefont{Liu}}, \bibnamefont{and}
  \bibinfo{author}{\bibfnamefont{J.}~\bibnamefont{Sapirstein}},
  \bibinfo{journal}{Phys. Rev. A} \textbf{\bibinfo{volume}{51}},
  \bibinfo{pages}{1196} (\bibinfo{year}{1995}).

\bibitem[{\citenamefont{Bridges and Wiese}(1970)}]{neonexp}
\bibinfo{author}{\bibfnamefont{J.~M.} \bibnamefont{Bridges}} \bibnamefont{and}
  \bibinfo{author}{\bibfnamefont{W.~L.} \bibnamefont{Wiese}},
  \bibinfo{journal}{Phys. Rev. A} \textbf{\bibinfo{volume}{2}},
  \bibinfo{pages}{285} (\bibinfo{year}{1970}).

\bibitem[{\citenamefont{Johnson et~al.}(1995)\citenamefont{Johnson, Plante, and
  Sapirstein}}]{dan}
\bibinfo{author}{\bibfnamefont{W.~R.} \bibnamefont{Johnson}},
  \bibinfo{author}{\bibfnamefont{D.~R.} \bibnamefont{Plante}},
  \bibnamefont{and}
  \bibinfo{author}{\bibfnamefont{J.}~\bibnamefont{Sapirstein}}, in
  \emph{\bibinfo{booktitle}{Advances in Atomic and Molecular Physics}}, edited
  by \bibinfo{editor}{\bibfnamefont{D.}~\bibnamefont{Bates}} \bibnamefont{and}
  \bibinfo{editor}{\bibfnamefont{B.}~\bibnamefont{Bederson}}
  (\bibinfo{publisher}{Academic Press}, \bibinfo{address}{San Diego},
  \bibinfo{year}{1995}), vol.~\bibinfo{volume}{35}, p. \bibinfo{pages}{255}.

\end{thebibliography}

\end{document}